\documentclass[aps,showpacs,dvips,showkeys,preprint,preprintnumbers,nofootinbib,11pt]{revtex4}
\usepackage{graphicx}
\usepackage{color}
\usepackage{amsmath}
\usepackage{amssymb}
\usepackage{bm}
\usepackage{multirow}

\def \be  {\begin{equation}}
\def \ee  {\end{equation}}
\def \ee  {\end{equation}}
\def \bea {\begin{eqnarray}}
\def \eea {\end{eqnarray}}

\begin{document}

\preprint{ECTP-2019-02}    
\preprint{WLCAPP-2019-02}
\hspace{0.05cm}
\title{Equation of State for Cosmological Matter at and beyond QCD and Electroweak Eras}

\author{Abdel Nasser Tawfik}
\email{atawfik@nu.edu.eg}
\affiliation{\scriptsize Nile University, Egyptian Center for Theoretical Physics (ECTP), Juhayna Square of 26th-July-Corridor, 12588 Giza, Egypt}
\affiliation{\scriptsize World Laboratory for Cosmology And Particle Physics (WLCAPP), 11571 Cairo, Egypt}

\author{Igor Mishustin}
\affiliation{\scriptsize Frankfurt Institute for Advanced Studies, Ruth Moufang Str. 1, 60438 Frankfurt, Germany}
\affiliation{\scriptsize National Research Center, Kurchatov Institute, 123182 Mocow, Russia}

\date{\today}

\begin{abstract} 

Various thermodynamic quantities for baryon-free matter are calculated by combining the most reliable non-perturbative and perturbative calculations, especially the most recent ones including as many quark flavors as possible. We extend these calculations by including other degrees of freedom (dof), such as photons, neutrinos, leptons, electroweak particles, and Higgs bosons, that allows us to consider the temperatures up to the TeV-scale. The calculations show that similar to QCD, the EW phase transition is also a crossover. We have found that while the equation of state for the hadronic  matter is linear, $p/\rho\simeq 0.2$, the one for higher temperatures is rather complex; it exhibits two crossover-type phase transitions, corresponding to strong and EW matter. At even larger energy densities, the deduced EoS becomes linear again and close to ideal gas. The combined equation of state can be used for modeling the expansion of the Universe from very early times and through the EW and QCD era.

\end{abstract}



\maketitle


\section{Introduction}

The standard model of cosmology potulates that the cosmic matter at large scale is homogeneously and isotropically distributed. Therefore, it can be  characterized by a barotropic equation of state (EoS), where the thermodynamic pressure depends on temperatures and various chemical potentials, only via the energy density. The energy density plays a fundamental role in the Einstein field equations \cite{Einstein1916} by determining the cosmological expansion rate of the Universe. For a uniform cosmic fluid, the Einstein equations reduce to the Friedmann equations, which can be solved, exactly, almost for any EoS without any limitation on the temporal variation. They can be applied for early as well as for late Universe. The time evolution of various cosmological parameters can be easily studied, if the EoS is correctly chosen for the various eras, through the history of the Universe. By using information from different high-energy experiments, e.g. the Higgs boson mass, recent lattice QCD simulations, and perturbative calculations, one can deduce  temperatures and densities, at different stages of the Universe expansion. In this paper, we shall consider the conditions corresponding to electroweak (EW) ($10^{-36}-10^{-10}~$s) and quantum chromodynamic (QCD) ($10^{-10}-10^{-3}~$s) phases after the Big Bang.

With this regard, it should be remarked that one of yet-unsettled problems of experimental particle physics is to directly measure the thermodynamic quantities, such as pressure and energy density of produced matter. In various high-energy experiments, the main challenge is to precisely measure the distribution of the produced hadrons and to distinguish them from each other. Apart from some electromagnetic probes, there is no direct access to the early partonic phase.  By the end of the day, one needs to apply various theoretical approaches to deduce the bulk properties of produced matter. They include different versions of the statistical models, lattice QCD simulations, and perturbative calculations. These methods enable us to construct an EoS, at extremely high temperatures, which are not yet reachable in high-energy experiments.

The present work is devoted to finding EoS for QCD and EW domains corresponging to ultrarelativistic temperatures and vanishing (or very small) chemical potentials, that might be used for different cosmological applications. To this end, we utilize the recent $2+1+1+1$ lattice QCD inputs for axion cosmology \cite{Berkowitz2015,FodorNature2016}. With a good control over all sources of systematic uncertainties, various thermodynamic quantities were calculated in lattice QCD for temperatures from a few MeV to $100$ GeV. Then, we have further extended this EoS by including photons, neutrinos, leptons, electroweak particles, and Higgs bosons, that allowed us to consider temperatures up to the TeV-scale. It was concluded in \cite{Laine2006} that relative to the free value, the Higgs dynamics slightly reduces the energy density at $T>160~$GeV. Also, it was confirmed that the charm quark seems to play a non-negligible role, even at lower temperatures. To this end, we utilize recent perturbative study on thermodynamics across the EW era deduced from lattice simulations using a dimensionally reduced effective field theory \cite{Laine2015} and then reconfirmed in more refind lattice QCD simulations \cite{DOnofrio20115}. The agreement between both sets of cacluclations is so excellent that we consider here only one of them; the earlier one \cite{Laine2015}.

Calculations of  the EoS by means of the lattice QCD with dynamical quarks and physical masses were carried out recently by several lattice groups. Determination of pressure, energy density, and effective dof as functions of $T$ represents a real challenge for particle physics and comology. Such a computational-time consuming approach dates back to $2006$, when QCD equation of state with $2+1$ staggered quark flavors has been deduced and the one-link stout action was greatly improved \cite{Fodor2006}. 

The rest of the present paper is organized as follows. A short review on the thermodynamic results obtained from the $2+1+1$ lattice QCD \cite{FodorNature2016} are described in Sect. \ref{sec:LQCDsims}. The perturbative calculations using recent lattice QCD simulations within dimensionally reduced effective field theory \cite{Laine2015}, are outlined in Sect. \ref{sec:pQCDsims}. These two sections give a fair credit to both non-perturbative and perturbative methods utilized earlier by several groups to obtain the basic thermodynamic quantities. This also opens the alternative way to include charm and bottom quarks. The combined results leading to extended EoS are presented in Sect. \ref{sec:rslts}. Conclusions are give in Sect. \ref{sec:concls}.

\section{Non-perturbative lattice QCD simulations}
\label{sec:LQCDsims}

\subsection{General Considerations}

Assuming that both quark masses $m_q$ and lattice spacing $a$ are dynamically depending on the guage coupling $\beta$, a four flavor staggered action with $4$ levels of stout smearing was utilized in the majority of recent lattice QCD simulations with as many as possible quark flavors see e.g. \cite{FodorNature2016}. Two combinations of $u$, $d$, $s$, $c$ flavor have been used; namely $2+1+1$ and $3+1$. It is obvious that both realizations are identical except for the masses of strange and charm quarks, whereas physical masses are assigned to $u$ and $d$ quarks ($m_{ud}=R\times m_s^{st}(\beta)$, where $1/R=27.63$, $\beta$ is the guage coupling, and $m_s^{st}(\beta)$ is the mass of $s$ quark). Although degenerate masses are assumed to the light quarks, a small isospin asymmetry was included analytically \cite{FodorNature2016}. Also, the mass of the $c$ quark is given as function of the gauge coupling $m_c=C\times m_s^{st}(\beta)$, with $C=11.85$. The temperature $T$ can be determined as a function of the temporal lattice dimension, $T=(a N_{\tau})^{-1}$. Alternatively, varying the gauge coupling $\beta$ leads to changing $T$ as well, although the spacial and temporal dimensions do not. The gauge coupling can not only allow for varying $T$, but also measures the pseudoscalar pion mass $m_{\pi}$ and the Wilson-flow based scale $\omega_0$, where $\omega_0=0.153\pm0.001~$fm and $m_{\pi}=712\pm5~$MeV. At $T=0$, $\omega_0$ gives the inverse flow time \cite{S19}.

After applying a Wilson-flow equation, the clover definition of the topological charge was made in $2+1+1$ and $3+1$ ensembles. To make the computational process more economic, an adaptive step size integration scheme was utilized. The time flow $(8 T^2)^{-1}$ was introduced in order to estimate the finite $T$ of both ensembles, where a variation in the time flow was allowed. This procedure has greatly contributed to reducing the systematic errors. To control their simulations, the authors of  Ref. \cite{FodorNature2016} have checked that these configurations lead to saturated susceptibility at large flow times \cite{FodorNature2016}. For the topological susceptibility, the topological charge with and without rounding has been used.

As already outlined for the $2+1+1$ simulations, same configurations have been used also for $3+1$ simulations. Here, the mass ratio of the charm quark and the degenerate lighter quarks  ($u$, $d$, and $s$) was taken to be $11.85$. As mentioned, for the masses of up- and down-quarks, the physical values are used, while the mass of the strange quark is a function of the guage coupling, $m_s^{\mathtt{st}}(\beta)$. At $T=0$,  simulations were done on $64 \times 32^3$ lattice with seven values of the lattice spacing descendingly ranging from $0.15$ to $0.06~$fm. But at finite $T$, the same parameters as in $2+1+1$ case were used. The topological charge was measured for every Hybrid Monte-Carlo trajectory. The configurations leading to a topology change were rejected. In other words, configurations were generated, at fixed topology. Quantitatively, this leads to an acceptance probability of about $40\%$ for the coarsest lattice, but higher probabilities for the finest ones.  

In the following section we review how various thermodynamic quantities were determined, non-perturbatively and perturbatively. Up to four quark flavors were included in the former case, but up to five quark flavors in the latter case.

\subsection{Lattice QCD equation of state in non-perturbative regime}

Adding up the contribution of the charm quark was done by several authors, for instance, by means of lattice QCD simulations \cite{S24,S63}, where the masses of some quarks were taken very heavy, i.e. they were partially quenched. Other lattice simulations have assumed that the four quarks are non-degenerate but have unphysical masses \cite{S23,S10}. In addition to these attempts, perturbative calculations have been also conducted in \cite{Laine2006}. Surely, the ultimate goal is to curry out non-perturbative lattice simulations with four dynamical quarks having the corresponding physical masses \cite{S21}. To this end, $2+1+1$ simulations with staggered action, and $4$ levels of stout smearing have been carried out in ref. {\cite{S11}. 

Calculating the lattice EoS means deducing the temperature dependence of pressure, energy density, and entropy from the trace anomaly, 
\bea
\frac{I(T)}{T^4} &=& \frac{\rho - 3 p}{T^4}, \label{eq:IT1}
\eea
which at vanishing chemical potential combines three quantities, $T$, $\rho$, and $p$. The entropy density is expressed as 
\bea
\frac{s}{T^3} &=& \frac{\rho + p}{T^4}.  \label{eq:TH1}
\eea
However, a direct estimation of pressure as function of energy density seems not feasible. Another restriction, which should be taken into consideration is that the lattice QCD simulations assume a universal thermal equilibrium. This assumption may not be true, if strongly interacting matter is undergoing a drastic phase transformation, or even a fast crossover. In this case one should expect non-equilibrium effects due to symmetry changing, etc.

The temperature independent divergence in the trace anomaly can be removed and its physical values can be evaluated accurately, when vanishing-$T$ ensemble is subtracted from each finite-$T$ one. Such a procedure seems to be not valid at high temperatures, which correspond to very fine lattice spacings. It is well-known that reducing the lattice spacing likely leads to increasing autocorrelation times, that in turn largely increases the computational costs. Alternatively, for each finite-$T$ ensemble one can generate a renormalization ensemble at exactly half of its temperature \cite{FodorNature2016}. Then, the physical trace anomaly can be obtained from the difference between each finite-$T$ and its half-$T$ ensembles, $[I(T)-I(T/2)]/T^4$, which is no longer divergent. Then, the resulting trace anomaly can be extended to lower temperatures and the dimensionless pressure can be deduced from the temperature integral of $I(T)/T^5$ \cite{FodorNature2016}. Finally, both energy density and entropy density can be estimated from Eqs. (\ref{eq:IT1}) and (\ref{eq:TH1}) as $\rho=I+3p$ and $sT=\rho + p$.

The section that follows elaborates how is the $5$-th quark flavor treated, perturbatively. This was achieved in several steps of tree-level corrections.

\subsection{Lattice QCD equation of state in perturbative regime}
\label{sec:lqcdpert}

It was found that the free energy calculated in Hard Thermal Loop (HTL) perturbation theory up to next-to-next-to-leading order (NNLO) \cite{S1}  agrees fairly well with the non-perturbative lattice QCD simulations for $2+1$ and $2+1+1$ theories. In drawing such a conclusion, it should be highlighted that the charm quark was treated as massless in the perturbative calculations but almost the physical mass was assigned to it in the non-perturbative ones. The charm quark mass threshold that enters the thermodynamical quantities, especially the equation of state, was first estimated perturbatively in \cite{Laine2006}. Then, it has been  estimated in a non-perturbative approach such as \cite{FodorNature2016}. In the perturbative calculations, the effects of the heavy quarks have been estimated to a lower leading order. Accordingly, the conclusion was made that this refers to a pressure ratio between $3+1$ and $3$ quark flavors QCD. Nevertheless, when comparing the pressure with and without the inclusion of the charm quark in both perturbative and non-perturbative calculations, a fair agreement has been obtained; less than $3\%$ \cite{FodorNature2016}. The tree-level correction due to the charm quark was suggested in the form 
\bea
\frac{p^{(2+1+1)}(T)}{p^{(2+1)}(T)} &=& \frac{SB^{(3)}(T)+F_Q(m_c/T)}{SB^{(3)}(T)}, \label{eq:chrm1}
\eea
where $SB$ stand for Stefan-Boltzmann approximations and $F_Q(m_c/T$ is the free energy density of a free quark with mass $m_c=1.29~$GeV.

The success with the inclusion of the charm quark encouraged to follow the same line with the bottom quark, i.e. to rewrite tree-level correction as in Eq. (\ref{eq:chrm1}). Having non-perturbative lattice QCD simulations for $2+1+1$ theory allowed to estimate the accuracy of this procedure. It is expected that the perturbative contributions dominate, especially at high temperatures ($\gtrsim 500~$MeV). In ref.\cite{FodorNature2016}, perturbative calculations have been performed up to $\mathcal{O}(g^6 \log g)$, see also refs \cite{S65,Brambilla}. 

Based on the good agreement of the perturbative pressure and the trace anomaly for $2+1+1$ quark flavors with the non-perturbative calculations at temperatures up to $1~$GeV, a continuation towards higher temperatures with the inclusion of the bottom quark seems to be eligible. A tree-level correction for the bottom quarks similar to that for the charm quark, Eq. (\ref{eq:chrm1}), was implemented, too. It was found that the ratio of the massless $2+1+1$ and $2+1+1+1$ perturbative pressure excellently agrees up to $0.3\%$ with the ratio in the SB limit. It was concluded that this tree-level correction works for heavier quarks, as well \cite{FodorNature2016}. Accordingly, reliable non-perturbative simulations at temperatures ranging between $500~$MeV and $10~$GeV seems to be feasible through the phenomenological approach, like  the one given in Eq. (\ref{eq:chrm1}). Such a wide $T$-range obviously covers various eras in the early Universe, where different phase transitions and, accordingly, different forces and degrees of freedom are involved. 

\section{Perturbative Calculations up to TeV Temperatures}
\label{sec:pQCDsims}

\subsection{Preliminary Remarks}

Over the years, the perturbative approaches are widely used for calculating cosmological EoS \cite{Laine2006,Laine2015,Laine2014}. At very high temperatures ($T\approx 160~$GeV), it is expected that a thermal EW phase transition (first order or crossover) should take place. Various thermodynamic quantities have been estimated across the EW phase transition {\it crossover}, see e.g. ref. \cite{Laine2014}. One should also mention studies of the EW baryogenesis, which should be consistent with Sakharov's conditions for generating the baryon asymmetry. Besides the baryon number non-conservation, this requires also the departure from the thermal equilibrium. The EW phase transition in rapidly expanding Universe, especially if it is of first order, should be followed by a drop in the temperature, for instance, due to the formation of a metastable phase. It is worth mentioning that an EW phase transition of first-order is expected in various  extensions of the standard model for the elementary particles. For the sake of completeness, we refer to recent perturbative calculations \cite{FodorNature2016} leading to a slow crossover rather than a first order transition.

By combining perturbation results up to a largest leading order with the non-perturbative QCD calculations, Sec. \ref{sec:LQCDsims}, one can obtain a reasonable EoS for cosmological applications \cite{Laine2015,Laine2014}. This is conjectured to cover a very wide range of temperatures (up to $\sim 200~$TeV) including both electroweak and strong (QCD)  domains of cosmic matter.

\subsection{QCD Domain}

In this domain ($0.2\lesssim T\lesssim 1~$GeV)}, the four light quarks and the gluons are taken into consideration as building blocks of strongly interacting matter \cite{Laine2015}. The perturbative corrections to the ideal (noninteracting) Stefan-Boltzmana (SB) EoS can be determined up to different orders of strong coupling constant $\mathcal{O}(g)$. As already mentioned in section \ref{sec:lqcdpert}, the contributions up to $\mathcal{O}(g^6 \log g)$ are well known \cite{g6log1,g6log2,g6log3}, where $g$ is expressed as function of $N_c$ (number of colors), $N_f$ (number of massless quark flavors), $\mu_f$ (quark chemical potential), and $T$ (temperatures up to the EW scale).  As discussed in Ref. \cite{Laine2006}, though the extensive calculations have been done for higher orders in the perturbative expression, very little is known about its basic integrals. Only terms $\mathcal{O}(g^2)$, at $T=0$ and $\mu_f\neq 0$ have been analyzed. 

Concretely, taking into account contributions of the gluons up to $\mathcal{O}(g^6 \log g)$ and non-vanishing quark masses up to NLO $\mathcal{O}(g^2)$, the conclusion was made that the effect of quark masses seem to be very similar compared to that of LO $\mathcal{O}(g^0)$ \cite{Laine2006}. In particular, it should be mentioned that there is no explicit $\mathcal{O}(g^6 \log g)$ computation for finite quark masses. The authors of ref. \cite{Laine2006} have suggested an alternative procedure: start with $N_f=0$ and $N_c=3$ corresponding to very heavy quarks, then calculate the change in the pressure by lowering masses towards their physical values. As expected, the thermal pressure increases with decreasing quark masses. To summarize, while the lattice QCD simulations \cite{FodorNature2016} first determine the interaction measure (trace anomaly) $I(T,\mu_f,\cdots)$, see Eq. (\ref{eq:IT1}), from which other thermodynamic quantities can be derived, the perturbative calculations  start with the grand-canonical free energy density or the pressure normalized at vanishing $T$, where both quantities should vanish. Based on this renormalization process, any ultraviolet divergences are conjectured to be entirely removed. 

To proceed with the perturbative calculations towards the EW domain, existing lattice QCD simulations within the dimensionally reduced effective field theory, different successive steps should be completed. Firstly, the so-called {\it hard modes} should be removed through integration with respect to momenta or, alternatively, summations over the Matsubara frequencies ($2 \pi T$, $g T$, $g^2 T$, $\cdots$). Secondly, the effective mass parameters and the gauge couplings should be specified and normalized in the $\overline{\mathtt{MS}}$-scheme \cite{Laine2006}. Thirdly, the various connection factors should be estimated for changing temperature, at $\Lambda_{\overline{\mathtt{MS}}}$ fixed \cite{Laine2006}. 
These are obtained when multiplying the lattice results for $N_f=0$, $N_c=3$ by the ones obtained in the SB limit, see Eq. (\ref{eq:chrm1}). The connection factors facilitate the inclusion of quarks, at $T>200~$MeV.

\subsection{EW Domain}

With the assumption that both bottom and top quarks interact, weakly, and the Higgs potential parameters $v^2(\bar{\mu})$ and $\lambda(\bar{\mu})$ are functions of the normalization scale $\bar{\mu}$, the free energy density, at finite $T$ and $\bm{\mu}$, was expressed as \cite{Laine2006}
\bea
f(\nu, T, \bm{\mu}) &=& -\frac{1}{2} \nu^2(\bar{\mu}) v^2 + \frac{1}{4} \lambda(\bar{\mu}) v^4 + \sum_i^{n_q+n_g} \pm {J}_i (m_i(v),T,\mu_i), \label{eq:EWfenergy}
\eea  
where $v$ is Higgs expectation value, $m_i(v)$ is the tree-level mass of $i$-th particle,  ${J}_i (m_i(v),T,\mu_i)$ gives contributions of the physical dof (scalars, vectors, and fermions) and $\pm$ stand for bosons and fermions, respectively. $n_q\; (n_g)$ are number of quarks (gluons). 

Although the electroweak theory has various scales covering a wide range of temperatures and chemical potentials, following ref. \cite{Laine2006} we fix the normalization scale $\bar{\mu}$ to $100~$GeV. The electroweak contributions to the free energy density and pressure, Eq. (\ref{eq:EWfenergy}), are taken into account even at temperatures far beyond this scale. 

A few remarks on the phase transitions within SM is now in order. About two decades ago, it was conjectured that the SM should have a strong first order phase transition, while the reliable lattice QCD simulations repeatedly produce a smooth crossover \cite{Laine2006}. If this picture is indeed true, it wouldn't be possible to interpret the large-scale structure formation in the Universe \cite{peter2001}. Furthermore, the Sakharov's hypothesis, that the Universe has started with matter-antimatter symmetry, but acquired asymmetry at later stages by fulfilling at least one of his four conditions could not be realized. Moreover the EoS at EW-scale might be close to an ideal gas, as e.g. in ref. \cite{Laine2006}, while other studies show significant deviations \cite{Gynther2006}. The strength of the thermal EW phase transition depends on various SM Lagrangian parameters, which are difficult to estimate. With the recently discovered Higgs mass, the phase transition should be  a smooth crossover \cite{Kajantie1996,Aoki1999} and thus the electroweak baryogenesis wouldn't be possible.

\section{Construction of the Combined EoS}
\label{sec:rslts}

\subsection{Rescaling Procedure}

\begin{figure}[htb!]
\centering{ 
\includegraphics[width=8.cm,angle=-90]{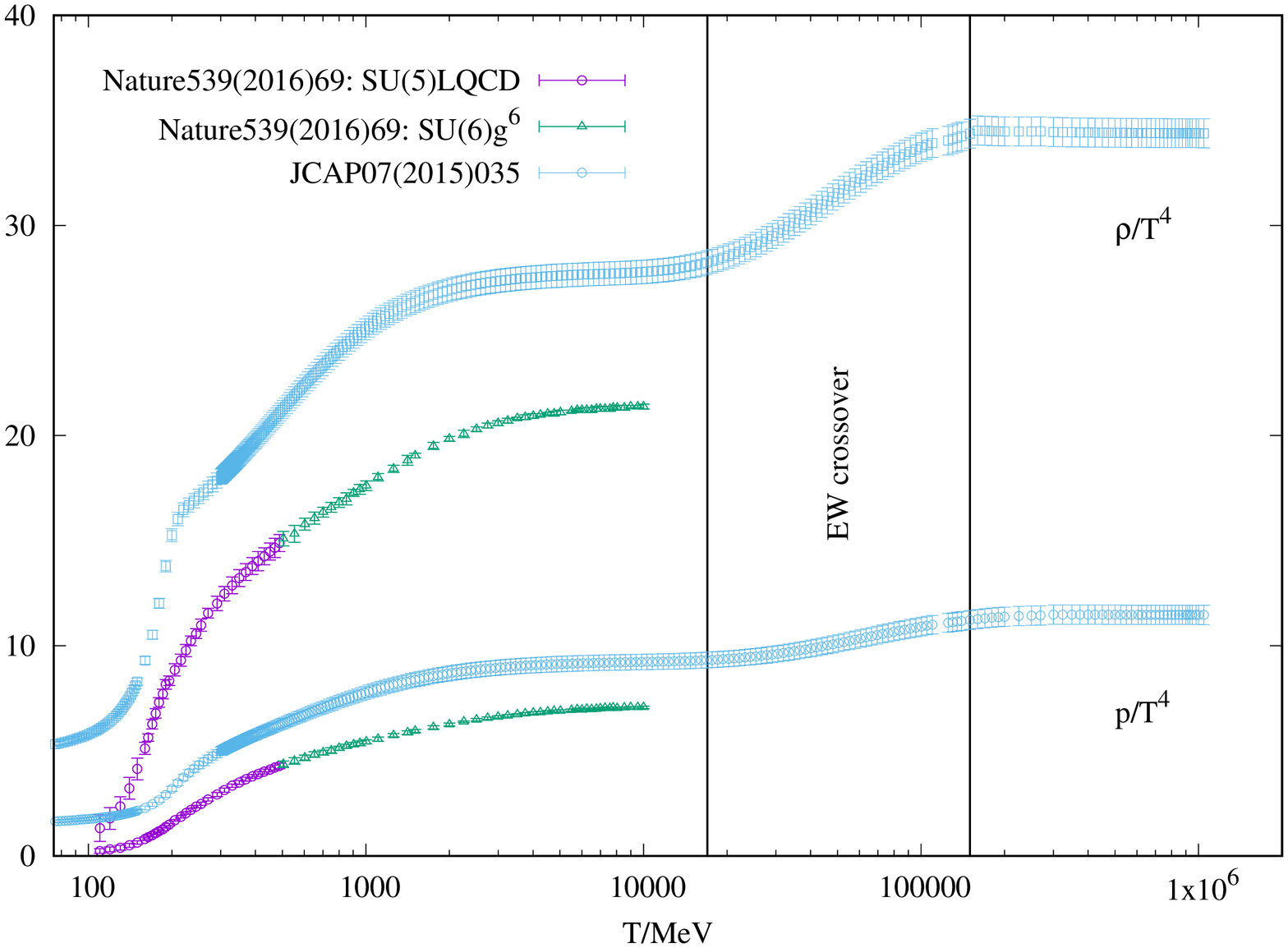} \\
\includegraphics[width=8.cm,angle=-90]{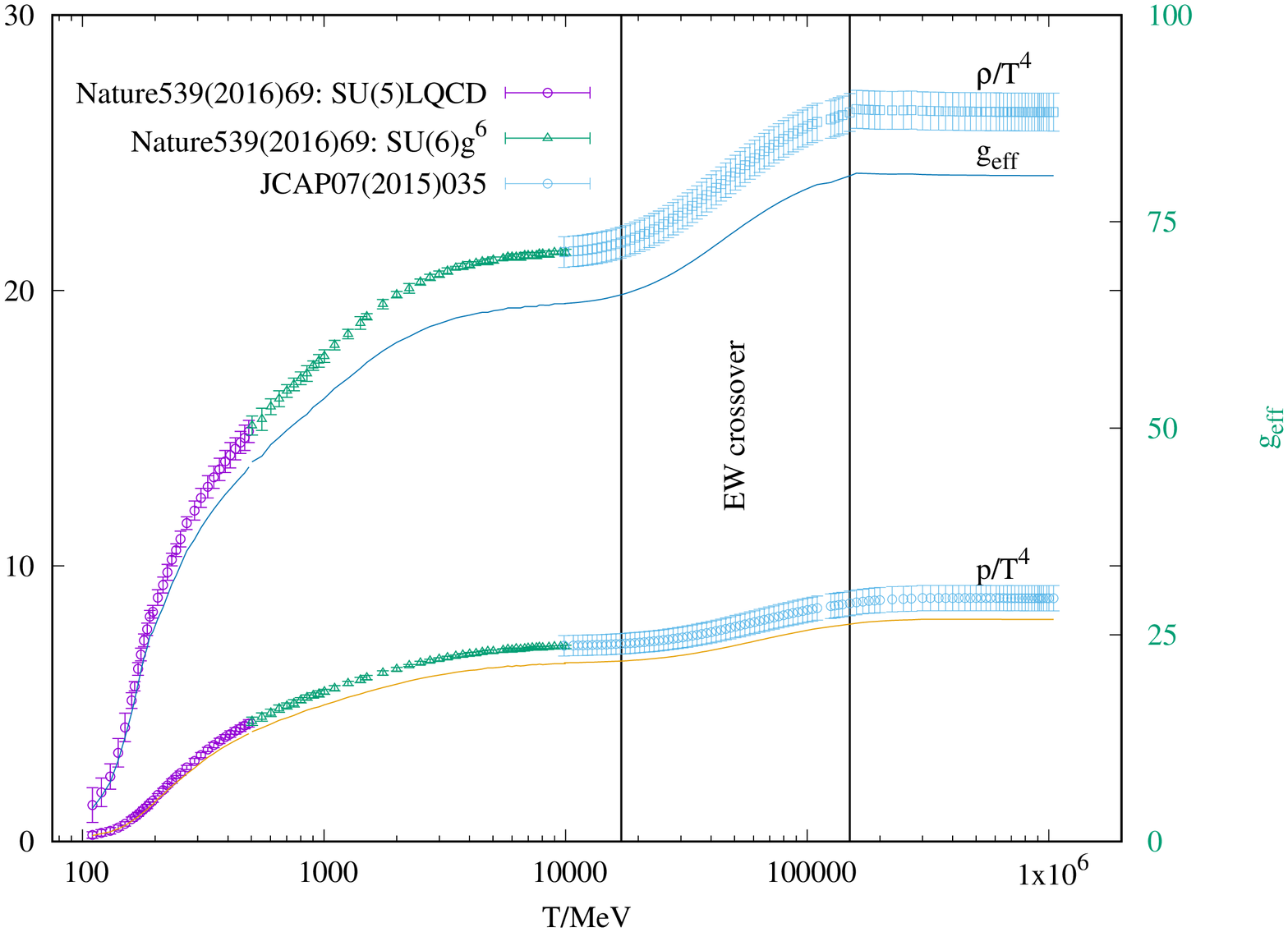} 
\caption{Top panel: a direct comparison between lattice QCD \cite{FodorNature2016} and perturbative calculations \cite{Laine2006} for normalized pressure and energy density as functions of temperature. Bottom panel shows the final results when both sets of simulations are properly matched (see text). The vertical lines roughly determine the temperature intervals, where the EW crossover transition is taking place. The curves in the bottom panel show the effective number of dof related to pressure and energy density. }
\label{fig:x}
}
\end{figure}

As  emphasized above, we intend to combine reliable non-perturbative  \cite{FodorNature2016} and perturbative \cite{Laine2006} calculations for various thermodynamic quantities, especially the most recent ones including as many quark flavors as possible. Then we add other dof, such as photons, leptons, and electroweak particles, and let the temperature go up to the TeV-scale, i.e. up to beyond the electroweak era. Accordingly, both sets of simulations unambiguously approach the physics of very early Universe. As a result, both standard models for the elementary particles and for cosmology seem to approach each other as never before.

The starting point, on which the present study is based, is illustrated in the top panel of Fig. \ref{fig:x}, where the normalized pressure and the normalized energy density of both sets of simulations are given as functions of temperatures. The temperature goes up to $\sim 10~$GeV and $\sim 1~$TeV in the non-perturbative and the perturbative simulations, respectively. It is obvious that both types of simulations are qualitatively similar, but quantitatively rather different. Namely, the perturbative results are significantly higher than the non-perturbative ones.

When combining both data sets, one should take into account that the region of validity of these two methods of calculations are different. The non-perturbative lattice QCD simulations are most relaible in the temperature range starting from the QCD scale, $T\gtrsim\Lambda_{\mathtt{QGP}}\approxeq 200~$MeV up to a few GeV. While the perturbative approach is reliable at higher temperatures up to TeV region. Then, to have a contineous behaviour, it would be quite reasonable to rescale the perturbative calcuations  within their region of validity \cite{Laine2006}. The rescaling should be adjusted to allow the perturbative results to match with the non-perturbative results at lower temperatures. To keep the systematic uncertainty as minimal as possible, we assume that such a rescaling factor ($0.77$) remains the same over all temperature range up to TeV or even beyond, as shown in the bottom panel of Fig. \ref{fig:x}. As the result, we obtain the thermodynamic quantities, which combine both non-perturbative ($T$ up to $\sim 10~$GeV) and perturbative  ($10~\mathtt{GeV}\lesssim T\lesssim 1~$TeV) calculations.  It is remarkable that both calculations, when properly rescalled, match each other very well. One can see that even the slopes of the two curves are the same at the matching point!. 

To normalize different thermodynamic quantities, for instance, energy density $\rho(T)$, an effective number of bosonic dof $\mathtt{g_{eff}}(T)=\rho(T)/\rho_0$, where $\rho_0=(\pi^2/30)T^4$ is the energy density for non-interacting gas of scalar massless bosons, was introduced in the exraction, see e.g. ref. \cite{Laine2006}. The effective dof $\mathtt{g_{eff}}(T)$ corresponding to pressure and energy density are depicted in bottom panel of Fig. \ref{fig:x}. It should be noted that in a dynamical system like the Early Universe one has to take into account that different particle species decouple from the thermal bath at different times. This will accordingly modify the number of effective dof.

\subsection{Analytical Parameterization of EoS}

\begin{figure}[htb!]
\centering{
\includegraphics[width=8.cm,angle=-90]{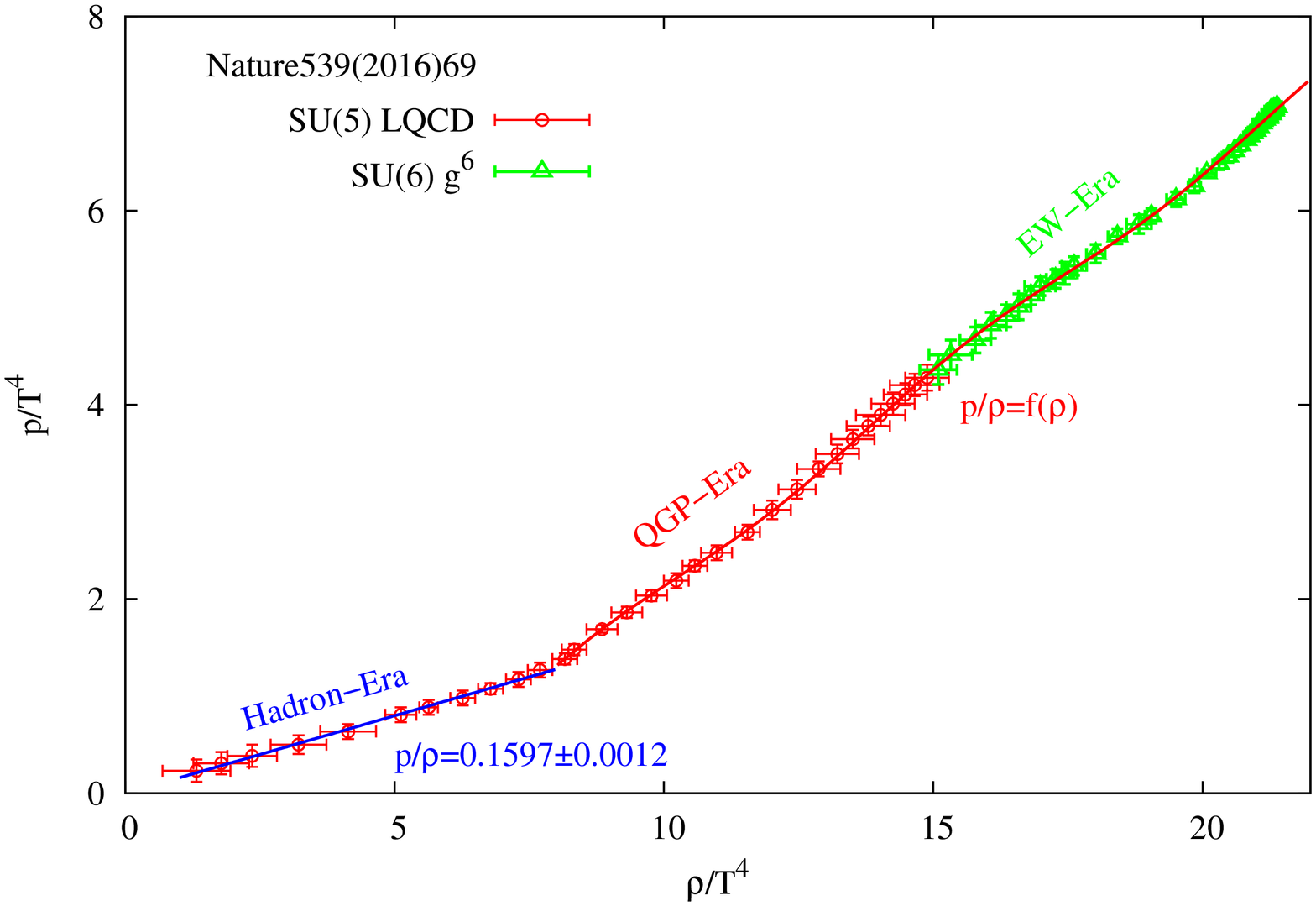} \\
\includegraphics[width=8.cm,angle=-90]{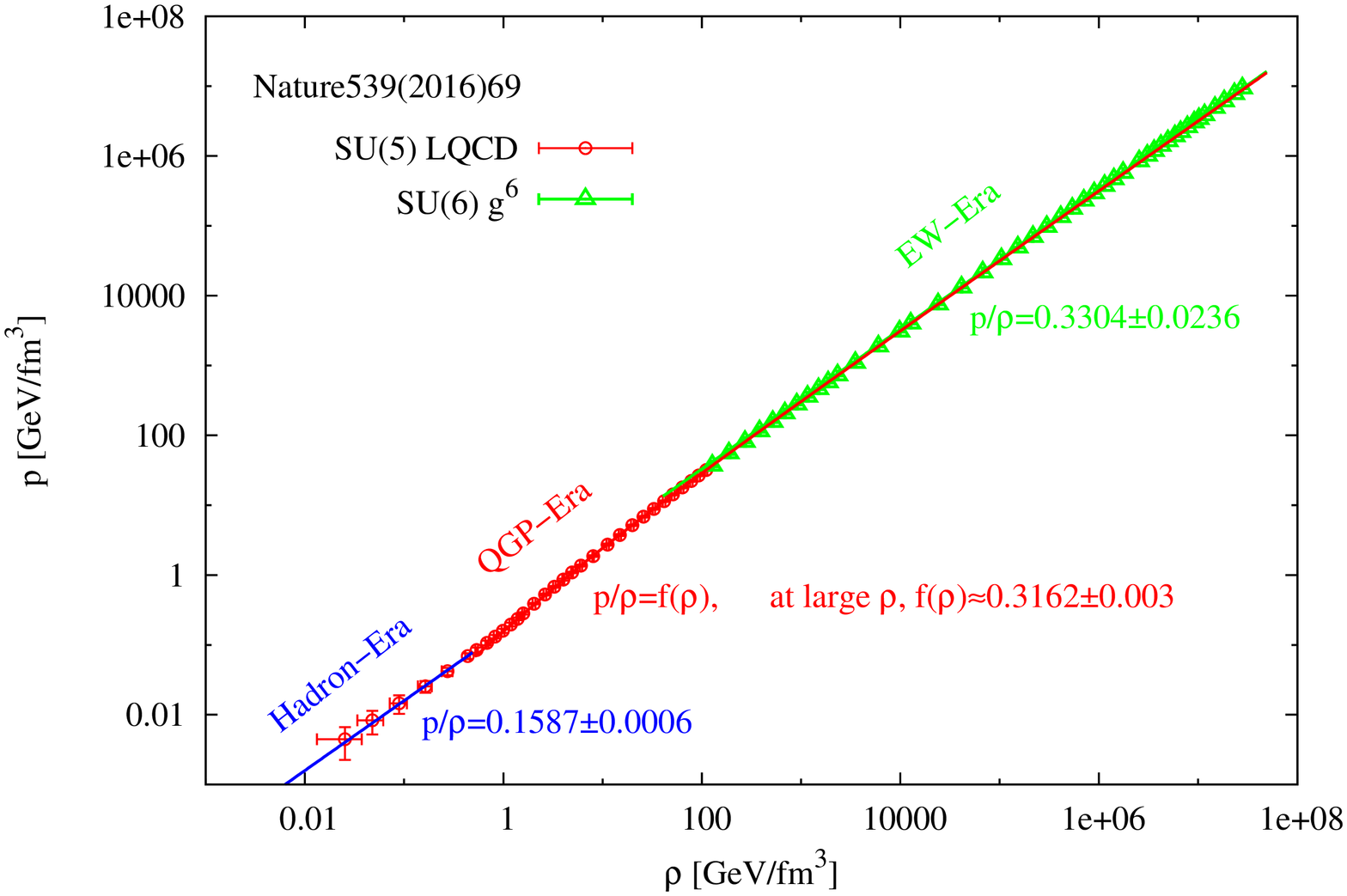} 
\caption{The normalized pressure is shown as function of the normalized energy density. Both quantities were calculated on lattice \cite{FodorNature2016} (top panel). The bottom panel shows the same but in the physical units on ($\rho - p$) plane.}
\label{fig:xx1}
}
\end{figure}

\begin{figure}[htb!]
\centering{
\includegraphics[width=8.cm,angle=-90]{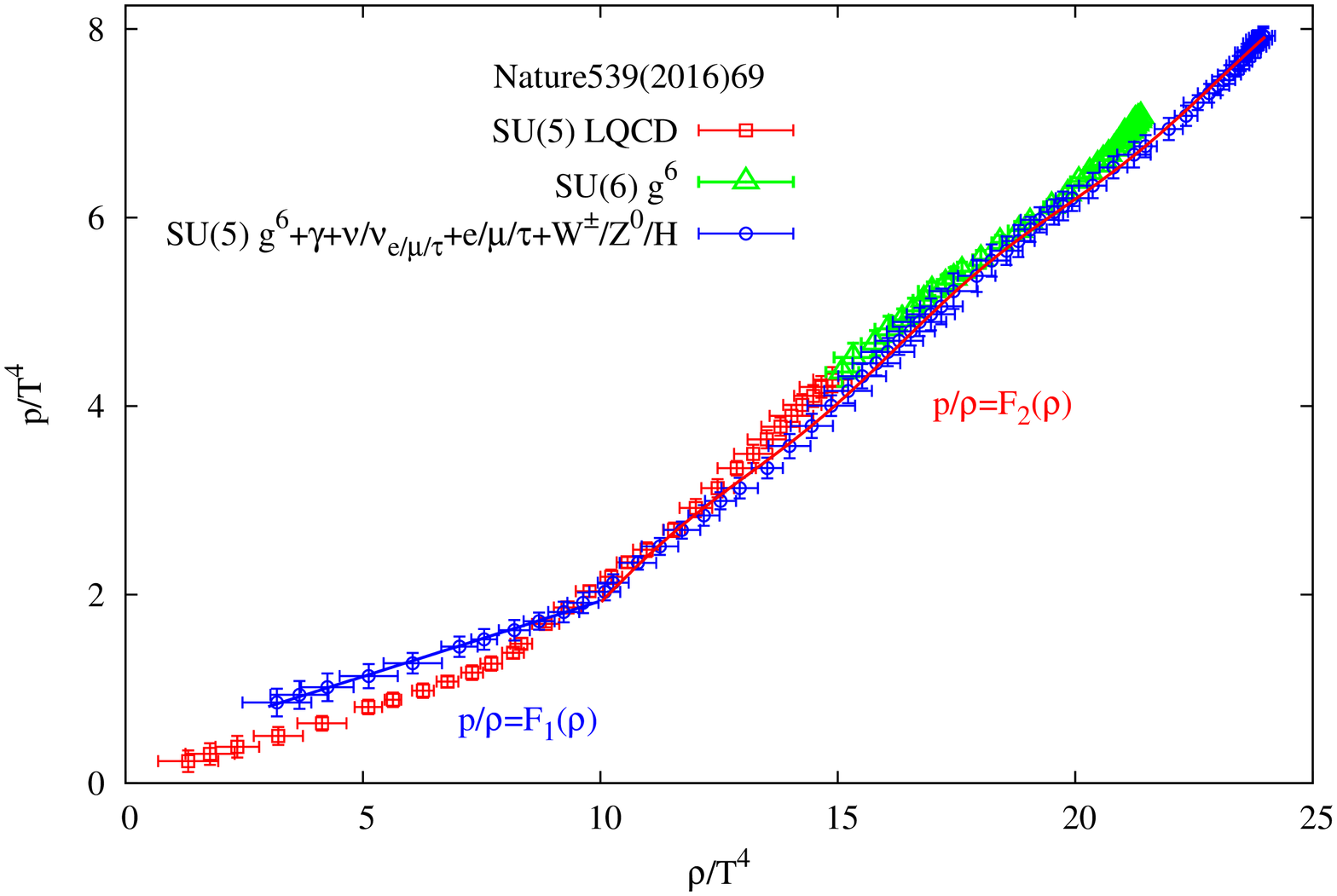} \\
\includegraphics[width=8.cm,angle=-90]{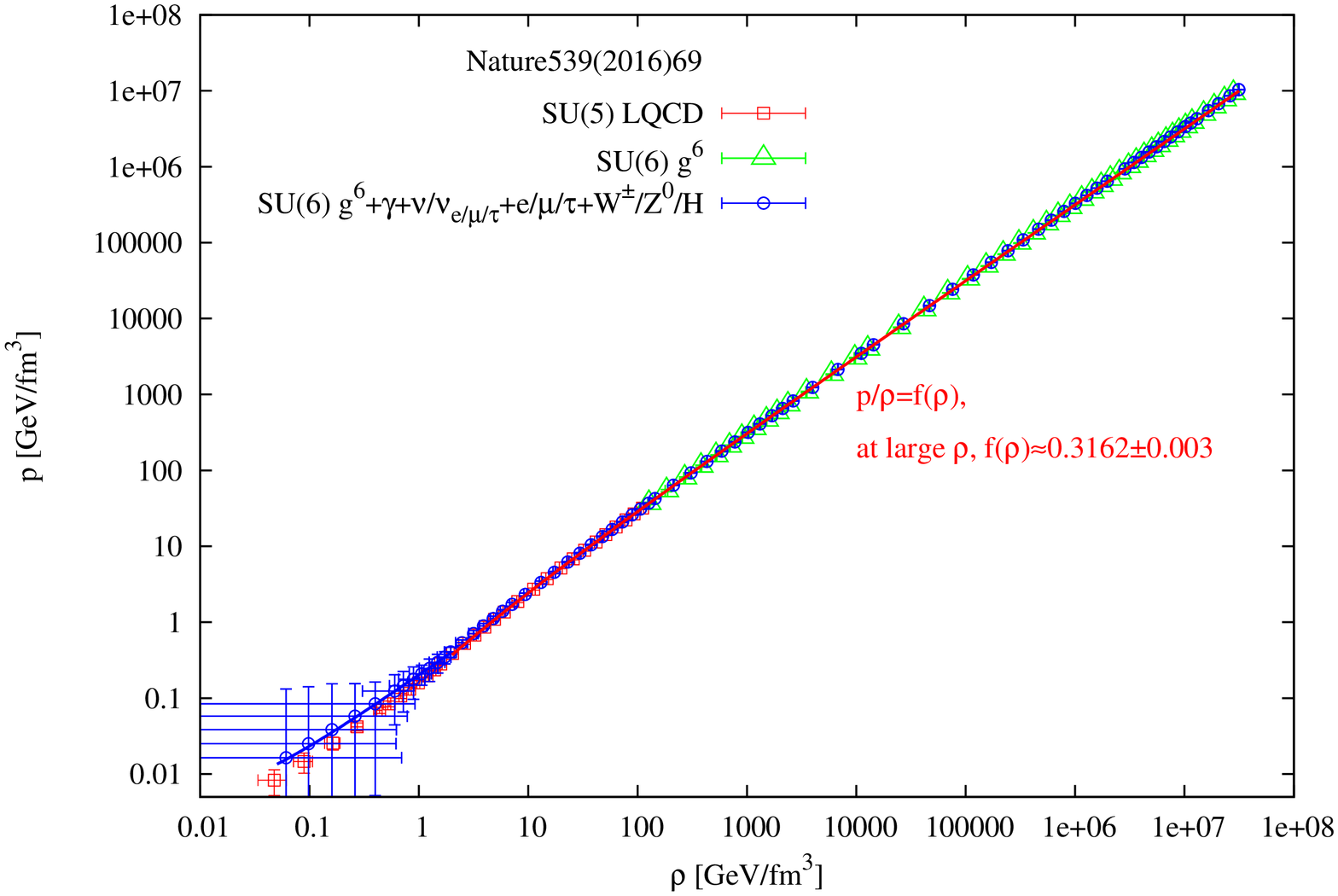}
\caption{The same as in Fig. \ref{fig:xx1} but both thermodynamic quantities were calculated on lattice \cite{FodorNature2016} and extended to include other dof such as $\gamma$, neutrinos, leptons, and EW and Higgs bosons (bottom panels). The bottom panel show the results in the physical units on ($\rho - p$) plane. Top panel compares between non-pertubative results with and without the inclusion of the additional dof.}
\label{fig:xx2}
}
\end{figure}

\begin{figure}[htb!]
\centering{
\includegraphics[width=8.cm,angle=-90]{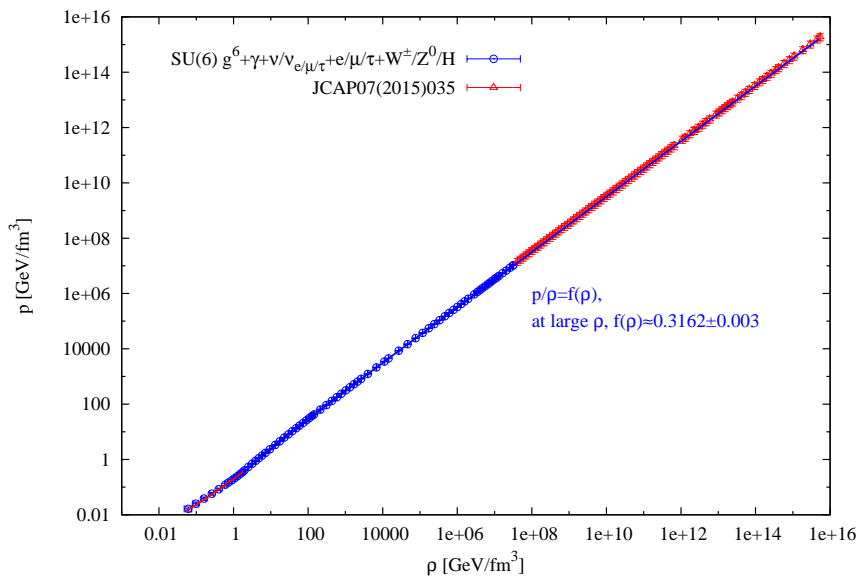}
\caption{The combined equation of state using non-perturnative \cite{FodorNature2016} with additional dof and perturbative results \cite{Laine2015,DOnofrio20115}.}
\label{fig:xxx}
}
\end{figure}

The contributions of the perturbative results to the combined EoS shall be depicted in Fig. \ref{fig:xxx}. Figures \ref{fig:xx1} and \ref{fig:xx2} show the EoS based on the non-perturbative lattice QCD simulations \cite{FodorNature2016}, where no scaling has been implemented. Figure \ref{fig:xx1} presents the non-perturbative lattice QCD simulations \cite{FodorNature2016} in a broad interval of energy densities up to $10^8~$GeV/fm$^3$, while Fig. \ref{fig:xx2} depicts the results when additionally other dof, such as $\gamma$, $\nu$, $e$, $\mu$, $\tau$, $\nu_e$, $\nu_{\mu}$, $\nu_{\tau}$, $W^{\pm}$, and the Higgs boson $H$ are added. It is obvious that these additional particles considerably contribute to each of both thermodynamic quantities. It seems that increasing the temperatures magnifies the contributions. On the other hand, the equation of state, i.e. the derivative of pressure with respect to the energy density, seems not being affected so much. A quantitative analysis will be presented, later.

As turned out, within the hadron phase (HP), the EoS is consistent with the linear dependence, $p/\rho=\mathtt{const.}$ However, the QGP and the EW phases can only be parameterized as a non-linear function.  We can parameterize those dependences (illustrated in the bottom panel of Fig. \ref{fig:xx1}) analytically as
\bea
\mathtt{HP:} \qquad  \qquad \frac{p}{T^4} &=& \alpha_1 \frac{\rho}{T^4}, \label{eq:eos11}\\
\mathtt{QGP/EW:} \qquad \frac{p}{T^4} &=& a_1 + b_1 \left[\frac{\rho}{T^4}\right]^{c_1} + d_1 \sin\left(\left[\frac{\rho}{T^4}\right]^{e_1}\right) \equiv f(\rho), \label{eq:eos12}
\eea
where the parameters are given in Table \ref{tab:1}.

\begin{center}
\begin{table}[htb]
\begin{tabular}{|c||l|}
 \hline
 \multirow{4}{4em}{$\mathtt{HP}$} & Eq. (\ref{eq:eos11}):\;\, $\alpha_1=0.1597\pm0.0012$  \\  
 &  Eq. (\ref{eq:eos21}):\;\, $\alpha_2=0.0034\pm0.0023$, $\beta_2=0.1991\pm0.0022$  \\
 &  Eq. (\ref{eq:eos41}): $\alpha=0.0034\pm0.0023$, $\beta=0.1991\pm0.0022$ \\\hline
\multirow{4}{4em}{$\mathtt{QGP/EW}$} &  Eq. (\ref{eq:eos12}):\;\, $a_1=-2.4456\pm0.1139$,  $b_1=0.4966\pm0.0312$, $c_1=0.9627\pm0.0167$, $d_1=0.08673\pm0.0058$, \\
 &\qquad\qquad\,  $e_1=0.9627\pm0.0014$ \\ 
& Eq. (\ref{eq:eos22}):\;\, $a_2=0.0484\pm0.0164$, $b_2=0.3162\pm0.0031$, $c_2=-0.21\pm0.014$, $d_2=0.576\pm0.034$ \\
& Eq. (\ref{eq:eos42}): $a=0.0484\pm0.0164$,  $b=0.3162\pm0.031$, $c=-0.21\pm0.014$, $d=-0.576\pm0.034$ \\ \hline
\multirow{2}{4em}{$\mathtt{Asymp.}$}  & Eq. (\ref{eq:eos23}): $\beta_2=0.3162\pm0.003$ \\   
& Eq. (\ref{eq:eos43}): $\gamma=0.3304\pm0.0236$ \\ \hline
\end{tabular}
\caption{Parameters from the varous fits for the EoS. All parameters are dimenionlss except $a_2$, $a$, $\alpha_2$, and $\alpha$. They have the physical units GeV/fm$^3$. \label{tab:1} }
\end{table}
\end{center}

Despite the non-simplicity of Eq. (\ref{eq:eos12}), we want to emphasize that this single EoS, $p/\rho=f(\rho)$, describes both EW and QCD matter, simultaneously. In our opinion this finding is rather interesting by itself and has a fundamental nature. Initially, we were interested in the realistic $p(\rho)$ for the strong matter, but it turns out possible to describe the EW matter, too. The two vertical lines in Fig. \ref{fig:x} roughly indicate the range of temperatures where the EW phase transition is taking place. Firstly, similar to the QCD phase transition, which was confirmed in various lattice QCD simulations, see for instance \cite{S10}, this is a crossover. Secondly, while the QCD phase transition takes place within the temperature interval of a few tens MeV, the EW phase transition extends over temperature range of a few tens GeV. In other words, the EW crossover is much smoother than that of QCD. In this paper we do not discuss possible consequences of these findings. 

To compare between the ($\rho-p$) results without and with the additional dof $\gamma$, $\nu$, $e$, $\mu$, $\tau$, $\nu_e$, $\nu_{\mu}$, $\nu_{\tau}$, $W^{\pm}$, and the Higgs boson $H$, both of them are drawn in the top panel of Fig. \ref{fig:xx2}. As mentioned, the bottom panel of Fig. \ref{fig:xx2} shows the same as the bottom panel of Fig. \ref{fig:xx1} but here we take into account the additional dof. By using the physical units GeV/fm$^3$, we hope to eliminate, as much as possible, the artificial temperature fluctuations. Also, expressing quantities in the physical units allows us better match of the SM of particle physics with the standard model of cosmology. Here, we arrive to the same conclusions, that a single function describes well both strong and EW domains, simultaneously. The fitting functions (illustrated in the bottom panel of Fig. \ref{fig:xx2}) are
\bea
\mathtt{HP:} \qquad \qquad p &=& \alpha_2 + \beta_2 \rho, \label{eq:eos21} \\
\mathtt{QGP/EW:} \qquad p &=& a_2 + b_2 \rho + c_2 \rho^{d_2}, \label{eq:eos22} \\
\mathtt{Asymp.:} \qquad p  &=& \beta_2 \rho, \label{eq:eos23}
\eea
where the parameters are given in Tab. \ref{tab:1}. We find that  at very large $\rho$ Eq. (\ref{eq:eos22}) can be precisely expressed as $p/\rho=0.3162\pm0.0031$, i.e. the first and the third terms of Eq. (\ref{eq:eos22}) almost cancel each other.

However, at asymptotically large $\rho/T^4$, the EoS becomes linear again, but slightly (of about $\lessapprox 2\%$) below the ideal gas limit, $\beta_1=0.3304\pm0.0236$. We conclude that $p/\rho=\mathtt{const.}$, at low $T$ and very high $T$, i.e. very simple,  while throughout the QCD and EW phase transitions, the behaviour is more complicated, $p/\rho=f(\rho)$, as given in Eq. (\ref{eq:eos12}).

To check the sensitivity of results to the choice of the numerical scheme, we have also performed calculations using refined lattice QCD of ref. \cite{FodorNature2016} combined with perturbative results based on dimensionally reduced effective field theory \cite{Laine2015}. This allows us to extend the energy density scale to $10^{16}~$GeV/fm$^3$. Figure \ref{fig:xxx} depicts pressure density as function of energy density in the physical units GeV/fm$^3$.

Again, we conclude that, for the hadron matter, the EoS is linear, i.e. $p/\rho=\mathtt{const.}$ and a single EoS describes well both EW and QCD domains. For the latter, the deduced EoS in the asymptotic limit becomes linear. Finally, we summarize the obtained results for EoS as:
\bea
\mathtt{Hadron:} \qquad p &=& \alpha + \beta \rho, \label{eq:eos41}\\
\mathtt{QGP/EW:} \qquad p &=& a + b \rho + c \rho^{d}, \label{eq:eos42} \\
\mathtt{Asymp.:} \qquad p  &=& \gamma \rho. \label{eq:eos43}
\eea
with parameters given in Table \ref{tab:1}.

The resulting EoS looks very similar to the one shown in Fig. \ref{fig:xxx} for lower energy densities. We believe that  Eqs. (\ref{eq:eos41}) and (\ref{eq:eos43}) represent the most reliable  EoS characterizing hadronic phase and QGP/EW matter, in a very broad domain of energy densities relevant for cosmological applications.

\section{Conclusions}
\label{sec:concls}

We have combined most recent non-perturbative and perturbative calculations for the thermodynamic pressure and energy density of baryon-free matter in the temperature range from about $100~$MeV to $1~$TeV. We have added the contributions of photons, neutrinos, leptons, electroweak particles, and Higgs bosons, and allow the temperatures go over the TeV-scale. In doing this, we have assumed that both non-perturbative and perturbative calculations, especially when they are properly  rescalled, should smoothly match with each other. Concretely, we have assumed  that the non-perturbative effects play an important role at $T\lessapprox 10~$GeV, when heavier quarks are also included in lattice simulations. At higher temperature, both pressure and energy density can be calculated, perturbatively, up to temperatures above $1~$TeV. 


We have found that the combined EoS is characterized by the number of effective dof; $\mathtt{g_{eff}}(T\sim1~\mathtt{TeV})\approx 80$, which is nearly $10\%$ below the value of non-interacting particles. We conclude that at low $T$ in hadronic phase, $p/\rho=0.1578\pm0.0006$. At very high $T$, the EoS is very close to ideal gas, $p/\rho=0.3304\pm0.0236$. But throughout the QCD and EW phase transitions, we have found that the EoS can be represented by a single function $p/\rho=f(\rho)$, which is extracted from the non-perturbative calculations. This might be useful for cosmological calculations. 

Our main conclusion is that having EoS from lattice calculations up to $T\sim 10~$GeV makes it possible to extrapolate it to higher temperatures using the perturbative approach as demonstrated in Figs. \ref{fig:x} and \ref{fig:xxx}. We have performed this analysis using two different sets of lattice simulations presented in refs. \cite{Laine2015,DOnofrio20115} and found that their predictions for the QCD and EW domains are fully consistent. Accordingly, both are crossover transitions. From the phenomenological point of view, it is important that the QCD phase transition turned out to be stronger than the thermal EW phase transition. We hope that out results will be useful to bring the standard model of elementary particles closer to the standard model for cosmology!

\section*{Acknowledgements}

The authors would like to thank Prof. Horst St\"ocker for supporting this project. The work of AT is supported by DAAD re-invitation funding program number $57378442$. AT is very grateful to Frankfurt Institute for Advanced Studies (FIAS) for hospitality and thanks Mikko Laine for providing parturbative calculations on the SM thermodynamic results from Ref. \cite{Laine2006}. IM acknowledges the financial support of the Helmholtz International Center for FAIR.



\end{document}